# Structure and Rheological Properties of Model Microemulsion Networks Filled with Nanoparticles

SHORT TITLE: FILLED MICROEMULSION NETWORKS

Nicolas Puech[1], Serge Mora[1], Vincent Testard[1], Grégoire Porte[1], Christian Ligoure[1], Isabelle Grillo[2], Ty Phou[1], Julian Oberdisse[1,3] *

[1] Laboratoire des Colloïdes, Verres, et Nanomatériaux (LCVN), Université Montpellier II, UMR CNRS 5587, 34095 Montpellier, France

[2] Institut Laue-Langevin (ILL), 6, rue Jules Horowitz, BP 156, 38042 Grenoble, France

[3] Laboratoire Léon Brillouin (LLB), CEA Saclay, UMR 12, 91191 Gif-sur-Yvette, France

revised version, 16$^{th}$ of January 2008

**Abstract :**

Model microemulsion networks of oil droplets stabilized by non ionic surfactant and telechelic polymer $C_{18}$-PEO(10k)-$C_{18}$ have been studied for two droplet-to-polymer size ratios. The rheological properties of the networks have been measured as a function of network connectivity and can be described in terms of simple percolation laws. The network structure has been characterised by Small Angle Neutron Scattering. A Reverse Monte Carlo approach is used to demonstrate the interplay of attraction and repulsion induced by the copolymer. These model networks are then used as matrix for the incorporation of silica nanoparticles (R=10nm), individual dispersion being checked by scattering. A strong impact on the rheological properties is found for silica volume fractions up to 9%.

**Figures:** 11

**Tables:** 1

(*) author for correspondance

**I. Introduction**

The structure, dynamics, and rheological properties of networks of telechelic polymers, i.e. triblock copolymers consisting of a middle block with 'stickers' at the extremities, have attracted considerable interest in the past decade. Model systems are usually oil-in-water or water-in-oil microemulsions bridged by the copolymer, which we shall call 'microemulsion networks' [1-12], and sometimes pure micelles formed by the copolymer [1, 13, 14]. Very recently, networks of connected wormlike micelles have also attracted interest [15, 16].

The scope of our work is to study the structure and rheology of networks of spherical microemulsion droplets filled with substantial quantities of inorganic nanoparticles, and a first approach is presented in this article. In the literature, particles in the nano- to micron-range have usually been added at very low concentrations as probe to networks, e.g. in order to obtain microrheological information [17]. In some studies, self-organised surfactant phases have been doped with concentrations of a few percent of particles, like inclusions of latex [18] or clay platelets [19, 20] in lamellar phases, in nematic (lyotropic) phases [21-23], inorganic colloids in columnar phases [24, 25], silica in wormlike micelles [26, 27] or copolymer gels [28]. Usually, the scope of these articles was to determine phase boundaries of thermodynamically stable mixtures, the changes in rheology induced by the particles, or the location of the (sometimes nano-, often micro-) particles. For instance, particles may be inside [24] or outside [25] the columns of hexagonal phases. An opposite approach has been sometimes chosen in the context of particle floculation, observing the effect of adding polymer to a silica solution [29].

In the present article, we study a model system of a microemulsion network inspired by the work by Michel et al [8]. A new system with octanol as a cosurfactant has been set up, and essential information on the phase behaviour of this new system is shown in section III.1. The rheological properties of microemulsion networks, and namely the percolation behaviour, are presented in section III.2. Moreover, two fundamentally new aspects not studied in the previously cited literature are addressed. First, the microemulsion droplet ratio is varied, which allows us to control the size ratio between the droplet and the telechelic polymer. This is of both fundamental and practical interest, because the size ratio should directly influence the proportion of bridges and loops, and thereby the colloidal pair potential induced by the polymer. Secondly, we introduce mineral particles, nanometric $SiO_2$ (R ≈ 10 nm), in the microemulsion network. We explore their effect on the rheology of the microemulsion networks in section III.3. Structural studies are presented in the second part of part III: In section III.4, Small Angle Neutron Scattering (SANS) is used to characterize the microemulsion networks, using a Reverse Monte Carlo (RMC) algorithm to convert data into real-space correlations. Finally, the structure and good dispersion of the silica in the microemulsion network is checked by scattering in section III.5, taking advantage of contrast variation by isotropic substitution. As mentioned above, the polymer-droplet size ratio is found to influence the colloidal pair potential. As illustration, some recent evidence for this dependence obtained by numerical simulations are briefly presented in section IV [30].

## II. Experimental

**Sample preparation:** Synthesis of the triblock copolymer $CH_3$-$(CH_2)_{17}$-NH-CO-O-$(CH_2$-$CH_2$-O$)_n$-CO-NH-$(CH_2)_{17}$-$CH_3$ has been described elsewhere [31-33]. Here we study a 10k-poly(ethylene oxide) chain which has a hydrophobic $C_{18}$-sticker at each extremity, connected through a urethane group. Triton X-100 (TX100), octanol, decane, were used as received. Bindzil silica (B30/220, 30%wt) suspended in water was a gift from Akzo Nobel. The pH of the delivered stock solutions was set between 9 and 10 in order to ensure maximal colloidal stability to the silica by electrostatic repulsion. All silica-containing samples have thus been prepared at pH 10, all others at pH 7, and we have checked that, e.g., rheological properties or scattering of unfilled connected microemulsions do not vary in this pH range. The temperature for all studies was 22°C, an important parameter given the sensitivity of non-ionic systems to temperature. The choice of a non-ionic surfactant is motivated by the absence of electrostatic interactions between the microemulsion and the colloidal silica, avoiding strong adsorption of a cationic surfactant on the negatively charged silica. Possible hydrogen bonding between PEO and anionic surfactants ruled out the latter. Samples are characterized by a volume fraction of microemulsion $\Phi$, which includes the surfactant, the cosurfactant and the decane. The cosurfactant-to-surfactant mass ratio is called $\Omega$, the oil-to-surfactant mass ratio $\Gamma$:

$$\Phi = \frac{V}{V+V_s} \qquad \Omega = \frac{m_{oct}}{m_{TX100}} \qquad \Gamma = \frac{m_{dec}}{m_{TX100}+m_{oct}} \qquad (1)$$

The volume V of oil and surfactants is estimated from the masses and their macroscopic densities ($d_{TX}$ = 1.07 g/cm$^3$; $d_{oct}$ = 0.83 g/cm$^3$; $d_{dec}$ = 0.73 g/cm$^3$; $d_{si}$ = 2.2 g/cm$^3$), and $V_s$ denotes the volume of the (aqueous) solvent. The amount of silica nanoparticles is expressed as the volume fraction of silica, using $\rho_{si}$ = 2.2 g/cm$^3$. This density is consistent with the scattering length density ($\rho_{si}$ = 3.5 10$^{10}$ cm$^{-2}$) determined by independent contrast variation experiments. Analogous experiments have been performed for with the microemulsion system and experimental scattering length densities are reported in Table 1.

The number of polymer molecules is given as the number of polymer stickers per droplet r, i.e. if there is on average one polymer molecule per droplet, r equals two. r is calculated from the weighted masses and the droplet radius as determined by SANS and given in Table 1.

The order of mixing of the components is of importance due to the high viscosity of the microemulsion networks. Microemulsions were prepared first, adding then the colloidal silica solution, and finally the telechelic polymer.

**Small Angle Neutron Scattering** (SANS) experiments have been performed on beamline PACE at LLB (Saclay) and on beamline D11 at ILL (Grenoble). The configurations were (1m and 4.68m at 6Å; 4.68m at 13 Å) at LLB, and (1.2m, 5m, 20m at 6 Å; 34m at 10Å) at ILL. Samples have been prepared in D$_2$O, small quantities of H$_2$O from the silica stock solution have been accounted for in the calculation of the solvent scattering length densities. Empty cell scattering has been subtracted, and detector efficiency has been corrected as usual with light water. Data were normalized either by an empty beam measurement (LLB), or by using beamline constants determined from standards (ILL). The incoherent background subtraction was performed by comparing to Porod scattering expected for smooth interfaces.

**Rheology:** An strain-controlled Ares-RFS rheometer was used for oscillatory shear experiments in cone-plane geometry (R = 15 mm, 0.02 rad, gap 50 μm). Temperature was set to T = 22°C. Storage and loss moduli G'(ω) and G"(ω) were measured typically from ω = 1 to 200 s$^{-1}$, in the linear domain (γ = 10%).

## III. Results

### III.1 Phase behaviour

**Phase diagram of pure microemulsion**

The phase diagram of pure microemulsions at low concentrations is rather insensitive to the exact value of the concentrations, and a cut at fixed total mass fraction of 5% is shown in Figure 1. The y-axis represents the cosurfactant-to-surfactant mass ratio $\Omega$, and the x-axis the oil-to-surfacant mass ratio $\Gamma$. Pure Triton X-100 micelles, usually described as ellipsoids, are found at the origin ($\Gamma = \Omega = 0$) [34]. As cosurfactant is added (along a line $\Gamma = 0$), a classical sequence in aggregate morphology is found [35]: Globular micelles, cylindrical micelles, and bilayers. With increasing $\Omega$, the spontaneous curvature of the hydrophobic-hydrophilic interface decreases, which induces micellar growth, yielding cylindrical (or wormlike) micelles. Above $\Omega \approx 0.15$, the system phase separates in a micellar phase and a lamellar phase favoured by the decreasing spontaneous curvature, and between $\Omega \approx 0.3$ and 0.4, a homogeneous lamellar phase ($L_\alpha$) is found.

If on the other hand the oil ratio $\Gamma$ is increased from zero ($\Omega = 0$), the oil is solubilized in the micelles, which first take a spherical shape, with an increasing size of the hydrophobic core. Naturally, bigger droplets have a less curved interface, and the energy penalty paid by the surfactant layer which can not adopt its spontaneous curvature will rapidly grow with size. At a certain point, the droplets can not grow any more and the oil phase separates, indicating emulsification failure ($\Gamma > 0.1$; $\Omega = 0$). The key point of this discussion is that increasing $\Omega$ leads to less curved interfaces, whereas increasing $\Gamma$ leads to bigger hydrophobic cores. Increasing these two parameters simultaneously should thus create bigger microemulsion droplets at thermal equilibrium.

In Figure 1, such a one-phase region ('the tunnel') is shown to exist in the phase diagram. It allows us to create different microemulsions, in order to investigate the effect of the droplet sizes. In Figure 2, the typical radius of the droplets (at 1%v) measured by SANS in the Guinier domain is shown as a function of $\Gamma$ (for different $\Omega$, all located in the 'tunnel'). Two intensities I(q) are also shown in the inset. They correspond to quite different radii, as evidenced from the concomitant increase of $I_o = I(q\rightarrow 0)$, the decrease of the Guinier domain at low q, and the decrease of the high-q Porod decay (not shown), indicative of a lower specific surface. We have also checked that at higher concentrations the droplets keep their size, as it can be estimated from the position of the structure factor peak. In summary, the droplet radius is found to increase considerably with $\Gamma$, namely by a factor of about three within our experimental range. In this study, we have chosen the two extreme points in the phase diagram (cf. Figure 1), representing two quite different droplet sizes: (A) Swollen micelles ($\Omega = 0$, $\Gamma = 0.05$, R = 32 Å), (B) Microemulsion droplets ($\Omega = 0.4$, $\Gamma = 0.7$, R = 104 Å). Parameters of system A and B are summarized in Table 1.

**Phase diagram of microemulsion networks:**

The next step is the formation of the microemulsion network by adding telechelic polymer. The triblock copolymer has a hydrosoluble central block of PEO (10k), and two hydrophobic $C_{18}$-blocks called 'stickers' because of their natural tendency to stick to the hydrophobic microemulsion droplets with an energy of approximately 20 kT. Such systems have been studied to great detail in the past [7-9, 36, 37], and we recall only the most important points. For a given droplet size, the quasi-ternary phase diagram is usually presented as a function of the droplet volume fraction, $\Phi$, and the number of stickers per droplet, r. The latter is a measure for the amount of added polymer, or network connectivity. The robustness of the generic shape of the phase diagram in these axes has been evidenced by M. Filali et al [36]. Qualitatively, a different behaviour is found for 'high' and 'low' $\Phi$, a distinction quantified by the ratio between the typical polymer size compared to the distance between droplets: At high $\Phi$, microemulsion droplets are close enough to be bridged by polymer. At low $\Phi$ and low r, droplets are isolated and both stickers of a polymer molecule stick to the same droplet, the so-called 'flower' conformation. At low r, the system is monophasic for all $\Phi$. At high r, typically above r = 6 or 8, the system forms isolated flowers at very low $\Phi$, then crosses a two-phase region where a dense and connected phase is separated from a dilute phase. The physical origin of the phase separation lies in the attraction induced by the bridging polymer, which gains in conformational entropy if both stickers can be in different droplets, a mechanism which compensates the loss in translational entropy of particles [38-40]. At high $\Phi$ finally, an entirely connected microemulsion phase is found. In this article, we study such microemulsion networks.

We have verified that the same behaviour is found in our system. System B, e.g., has been studied at a volume fraction of $\Phi = 20\%$. The microemulsion droplet radius being about 104 Å, one can estimate the typical surface-to-surface distance to 80 Å. This distance has been fixed for all our samples, by adjusting the concentration for each droplet size (A: 4%, B: 20%). The scope was to fix it to the same order of magnitude as the average end-to-end distance of the PEO, which is approximately 85 Å for the 10k. At the chosen concentrations, all samples between r = 0 and r = 20 are stable and monophasic. Although we have not redone the complete ($\Phi$,r) phase diagram, we have checked that a lower volume fraction ($\Phi = 10\%$) for system B, e.g., the system phase separates as expected at intermediate r (r = 10). For smaller droplet sizes, we have created connected microemulsions with system (A), and the concentrations and parameters are recalled in Table 1.

**III.2 Rheological properties of microemulsion networks**

The rheological properties of the microemulsion networks have been characterized by oscillatory shear experiments. The shear strain amplitude was set to $\gamma = 10\%$, which we have checked to be safely in the linear domain, and storage and loss moduli were determined as a function of $\omega$. As with the data presented in the literature, G' and G'' of microemulsion networks show essentially Maxwellian behaviour, i.e. they can be characterized by a modulus G and a single relaxation time $\tau$ [6, 7, 11, 14, 37]. Minor deviations from Maxwellian behaviour are discussed later in the article.

We start with the swollen micelles, system A, to which increasing amounts of telechelic polymer are added. The average number of stickers per micelle is varied from r = 0 (no polymer) to r = 10. Storage and loss moduli were measured over the full frequency range, and fitted with the Maxwell model, deducing G and $\tau$. G is plotted as a function of r in Figure 3.

A clear change in rheological behaviour of the microemulsion network between r = 2 and 3 is evidenced in the plot. Below a critical connectivity $r_c$, there is not enough polymer to connect entire regions of the sample, and only small clusters are formed. The resulting sample is liquid-like, with a viscosity ten times higher than the one of pure swollen micelles, but still very low with respect to the percolated network ($\eta_{pure} = 1.2$ mPa s, $\eta(r=1) \approx 15$ mPa s, $\eta(r=5) \approx 7$ Pa s). Below percolation, the measured modulus G is too low to be measurable. Above $r_c$, a percolation phenomenon occurs, and the samples show gel-like behaviour. As proposed in the literature [8, 9], we have fitted the dependence of G on r with a percolation law of the type $G = G_o [(r- r_c)/r_c]^\beta$, and obtain $G_o = 800\pm200$ Pa, $r_c = 2.5\pm0.2$ and $\beta = 1.6\pm0.1$. As discussed by Michel et al, the exact fit parameters must be taken with caution, which is why we give quite large error-bar estimations, compatible with the spread in our data.

The relaxation time $\tau$ follows a similar percolation law, $\tau = \tau_o [(r- r_c)/r_c]^\beta$, with $\tau_o = 0.0085\pm0.0015$ s, $r_c = 2.5\pm0.2$ and $\beta = 1.3\pm0.1$. In a second experiment, we have measured the zero-shear viscosity $\eta$ in a steady shear experiment, increasing the shear rate $\dot\gamma$ from 1 to 100 s$^{-1}$. The viscosity is found to be agree nicely with a similar law, $\eta = \eta_o [(r- r_c)/r_c]^\beta$, with $\eta_o = 7\pm1$ Pa s, $r_c = 2.5\pm0.2$ and $\beta = 2.9\pm0.1$, where $\eta_o = G_o\tau_o$, and the exponents add up. Finally, we have also performed several equivalent experiments in the time domain, measuring the relaxation modulus G(t) after a small step strain (in the linear domain), and found identical (within error bars) results for G and $\tau$.

The rheological properties of the network made of bigger microemulsion droplets (system B) as a function of polymer content (r = 0 to 20) have been measured in the same way as with system A, and the modulus G is shown in Figure 4.

The data are slightly scattered around a clear, increasing trend. The modulus could be fitted with a percolation law, as well as the relaxation time $\tau$: $G_o = 110\pm20$ Pa, $r_c = 3.5\pm0.2$ and $\beta = 1.6\pm0.1$; and $\tau_o = 5\pm1$ ms, $r_c = 3.5\pm0.2$, $\beta = 0.7\pm0.1$. Below $r_c$, the moduli were again unmeasurably low, and the viscosity much lower, about three times the one of the pure system B ($\eta_{pure} = 3.5$ mPa s, $\eta(r=2) \approx 10$ mPa s, $\eta(r=8) \approx 1.8$ Pa s). Above $r_c$, the Maxwell relationship between a steady (low-) shear experiment and oscillatory shear in the linear domain, $\eta = G \tau$ is again found to nicely reproduce the data: $\eta_o = G_o\tau_o$, and the exponents of G and $\tau$ add up to the one of $\eta$ ($\eta_o = 0.55\pm0.1$ Pa s, $r_c = 3.5\pm0.2$ and $\beta = 2.3\pm0.1$).

A point common to both series of samples (A and B, Figs. 4 and 5) is that the percolation law fits our observations reasonably well, and a linear increase of G with r is not found, even far above percolation (r >> $r_c$). Such a behaviour has already been noticed by Michel et al [8]. Such a linear increase would be suggested by a very simple model of all polymer chains evenly distributed over the sample. Based on the assumption of Gaussian chain statistics, the modulus of a network is then given in a first approximation by $G = \nu\, kT$, where $\nu$ is the number of active links per unit volume, which would be simply proportional to r in our systems: $G = r/2\, \Phi/V_d\, kT$, $V_d$ being the volume of a droplet. It is nonetheless instructive to evaluate G with this simple law, and we find about 3000 Pa for r = 5 (system A), and about 900 Pa for r = 10 (system B). Note that the ratio between the moduli can also be estimated from the change in concentration (4% to 20%), in r (5 to 10), and droplet size. The experimental values are a factor of three lower in both cases (ca. 1000±200 Pa, and 250±50 Pa, for A and B, respectively). This illustrates two key points in this system: (a) Not all polymer molecules constitute active links, as some are certainly in a flower conformation, i.e. do not connect two micelles. The relative proportion of loops (flowers) and bridges, which may not be constant with r, is completely unknown at present. Our result suggests that in this case only ≈ ⅓ of the chains contribute to the modulus, the proportion being slightly higher in the case of the smaller droplets (system A). (b) Due to the statistical spatial distribution of polymer molecules, the onset of percolation can not be described by a model built on a homogeneous distribution. In this context, it should be mentioned that heterogeneous distribution of polymer has been discussed in terms of 'super-bridges' in the literature [1, 41].

**III.3 Modulus of filled microemulsions and networks**

We have incorporated silica nanoparticles in the microemulsion networks characterized in the unfilled state in the preceding section, varying the silica volume fraction $\Phi_{si}$ from 0% to 9%. In analogy with the reinforcement of elastomers, adding a volume fraction of $\Phi_{si}$ of hard particles, leaves (1 - $\Phi_{si}$) to the microemulsion. In scattering, it is wise to leave the total volume fraction of microemulsion droplets unchanged (avoiding shifts in position), whereas it seems conceptually advantageous to leave the water-to-droplet ratio unchanged in rheology. This means that the matrix around particles remains unchanged, and implies that the volume fraction of microemulsion droplets decreases slightly as more silica is added. In absolute numbers, the effect is almost negligible, e.g. adding 5%v of silica to a 4%v microemulsion (system A) decreases the micellar volume fraction to 3.8%v.

As with pure microemulsion networks, frequency sweeps were performed to measure G' and G", and fitted with the Maxwell model to extract G and τ. The data and the fits are shown in Figure 5, where we compare an unfilled network to a filled one ($\Phi_{si}$ ≈ 6%), all other parameters remaining unchanged. It is found that silica enhances the deviations in G" from Maxwellian behaviour at high frequency, and slightly modifies the power-law of G' in the low-frequency domain. Similar observations have been communicated in the literature, cf. Fig. 20 of ref [7]. In spite of the deviations, it is clear from the plots that G and $\tau$ can still be used as trustworthy characteristics of the filled network.

The evolution of the storage modulus G deduced from the Maxwell-fit with silica volume fraction in filled microemulsion networks has been measured between $\Phi_{si}$ = 0 and $\Phi_{si}$ = 7% for system A. They are plotted in Figure 6a for two network connectivities, r = 5 and r = 10. Motivated by the analogy with polymer-filler composites, we have plotted the reduced storage modulus $G/G(\Phi_{si} = 0)$ in Figure 6b. This quantity, called the reinforcement factor in elastomer reinforcement [42, 43], expresses how much stronger the 'nanocomposite' is with respect to the pure matrix. In spite of some scatter of the data, it is found that the shear modulus increases in both the weaker (r=5) and the stronger (r=10) network. In the reduced presentation, it is found that the relative increase is about the same, and reaches more than a factor of two with addition of a few volume-percent of silica. A linear fit of the reduced modulus gives (1 + 18 $\Phi_{si}$). In the absence of any quantitative theory on reinforcement in such a system, we can compare this tendency only to the simplest and oldest hydrodynamic theory, namely the increase of the viscosity of a dilute colloidal suspension by (1 + 2.5 $\Phi_{si}$), as calculated by Einstein [44, 45]. Smallwood has shown that the same equation can be used for the modulus of an elastic matrix [45], and we suppose here that it can be applied in a first approximation to Maxwellian viscoelastic fluids, as $\eta$ = $G\tau$. The influence of the silica volume fraction on $\tau$ does not seem to follow any simple law, but stays weak. Moreover, one may note here that calculation up to higher order in the volume fraction exist [46], namely up the quadratic order (1 + 2.5 $\Phi_{si}$ + B $\Phi_{si}^2$). However, in our case, there is no curvature in the data, and in any event B would be several hundreds, which is many times higher than theoretical value of 14 (or 6.2, the value proposed by Batchelor). It is concluded that the observed increase is at least an order of magnitude higher as predicted by simple theories, which suggests that the silica has a specific reinforcement effect on the microemulsion network, possibly introducing additional network nodes, or reorganising the network.

A similar analysis has been performed with the bigger microemulsion droplets of system B. In Figure 7a, the storage modulus G of filled microemulsion networks as a function of silica volume fraction ($\Phi_{si} = 0$ and $\Phi_{si} = 9\%$) is reported for three network connectivities, r = 10, 15 and 20. The reduced storage modulus G/G($\Phi_{si} = 0$) is represented in Figure 7b.

The data in the Figures 7a and b are more scattered than with the micelles of system A. Nonetheless the increase in modulus with $\Phi_{si}$ is of the same order of magnitude as before, and we find 1 + (11±4) $\Phi_{si}$, with a some large relative error due to the spread of the data, and namely the 'dip' at low $\Phi_{si}$. It is unclear at the moment, if this slight decrease observed at low silica volume fraction and low connectivity is due to a problem with the samples; the kinetics of network formation may be very slow due to the high viscosity of the samples. In spite of the large error bars, it is evident that the reinforcement effect is again considerably higher than simple hydrodynamic reinforcement, even if one takes the weak $\Phi_{si}$–dependence of $\tau$ into account, which is roughly proportional to (1 – 2$\Phi_{si}$) for r = 10 and r = 15.

**III.4. Structure of microemulsion networks:**

Investigations of the structure of microemulsion networks have been published in the literature [3, 5, 7, 37]. Data are sometimes interpreted by using ad-hoc potentials, as it is the case of a recent study at low volume fractions [47]. In general, it may be noted that microemulsion networks are rather special because the polymer molecules at the origin of the network elasticity are not easily visible by scattering techniques, like it would be the case for pure polymer networks. On the contrary, the nodes made of dense microemulsion droplets are highly visible.

In this article, the scattered intensities are briefly revisited and analysed in a new manner, based on a Reverse Monte Carlo analysis. The motivation for our scattering experiment goes beyond the understanding of pure networks, as we also wish to characterize the dispersion of the silica nanoparticles in the networks made of different droplets. First results on this topic are presented in section III.5. The effect of nanoparticles on the network will be studied in a forthcoming article, and we hope that this will lead to a deeper understanding of the puzzling rheological properties of the filled microemulsion networks. One may also note, e.g., that the increase in modulus of wormlike micellar solutions with micron-sized particles is considerably smaller [26].

In Figure 8, the evolution with connectivity r of the structure of system A is illustrated through the scattered intensities. All intensities show a slight upturn at small angles, which may be due to concentration fluctuations at large scales and which we neglect in the present discussion focussed on the local network structure. Pure micelles at 1%v (rescaled from Figure 2 to 4% for comparison) are found to have a nice Guinier-regime at small angles, yielding a radius of 32 Å. As the concentration is increased to 4%, the low-q intensity decreases, which is due to the influence of the hard sphere structure factor. Experimentally, we find a decrease to about 70%, slightly below the prediction of the Carnahan-Starling equation for hard spheres, S(0) = 0.73 [48]. As polymer is added, obtaining a connectivity of r = 5, the low-q intensity is suppressed even further, and a peak appears at $q_o = 0.0445$ Å$^{-1}$ (cf. arrow in Figure 8). This peak reflects the correlations inside the network of micelles, with a typical center-to-center distance of $2\pi/q_o = 140$ Å. If we relate this to the micellar volume using the volume fraction and a simple cubic lattice model, we find a micellar radius of 30 Å, i.e. a value consistent with the Guinier analysis. This is a strong indication that the peak is really due to nearest neighbour ordering in the liquid phase. Moreover, the typical surface-to-surface distance is about 75 Å, which is of the order of the polymer end-to-end distance, as desired. At large angles, finally, the three intensities are seen to superimpose, which means that there is the same amount of surface in these systems, and also that the polymer contribution is negligible with respect to the one of the micelles. It is concluded that the swollen micelles of system A do not change shape when building up the microemulsion network.

In the case of swollen micelles, it was possible to convert the structure as seen by scattering in direct space information by using a Reverse Monte Carlo (RMC) algorithm [49, 50]. Details of the calculation, which is based on concepts used for scattering from aggregates, are reported in the appendix [51]. The result is a three dimensional representation of droplet locations within a simulation box, the scattered intensity of which is identical to the measured one. The list of droplet locations can be used to extract the pair correlation function, or any other desired quantity. The RMC-algorithm is thus used to Fourier transform the intensity by forward modelling (from direct to reciprocal space), as it is well known that the inverse Fourier transformation is difficult if not impossible due to the loss of phase information, as well as insufficient sampling in q-space. Let us note that other ways from reciprocal to direct space exist, like e.g. the GIFT package [52-54].

In the inset of Figure 8, the pair correlation function for the same series (system A, Φ = 4%, r = 0, 3, 5) is shown. The corresponding scattering functions show almost no deviation from the experimental intensities. In order to accelerate the convergence of the system, we allow the droplets to violate the excluded volume condition, but expect collisions to be very rare because no such collisions can be present in the experimental system. Some collisions are nonetheless found, leading to non-vanishing values for the pair correlation function below two droplet radii. We have not represented them in the figure for clarity. In the inset of Figure 8, the system is seen to evolve from an almost structureless system at r = 0 with some short distance repulsion – the correlation hole just above the diameter – to a more organised system with a clear correlation peak at 140 Å for r = 5. This distance corresponds necessarily to the position of the peak in the scattered intensity, as already noted above. As the amount of polymer is increased, the peak hardly changes shape, position, or height, but the probability of presence at contact is found to decrease visibly. It is concluded that at r = 3, i.e. just above the percolation threshold, the system is already positionally correlated, and further increasing the amount of polymer only hardens the interaction, by adding to the entropic repulsion induced by the chains. In conclusion, adding further polymer molecules in between the micelles thus reduces the compressibility, which results in the decrease of the low-q intensity, and translates in real-space in a decreased probability of close-contact.

The bigger droplets (system B), display a quite similar behaviour, with two noticeable exceptions. First, the system is considerably more concentrated (20%v), which is due to the constraint of the surface-to-surface distance being comparable to the polymer size. Secondly, the droplets lose the signature of monodisperse spheres at large angles, i.e. oscillations visible in the Porod domain at 1%v disappear at 20%v. This makes it difficult to use the same RMC-approach as with system A, the form factor being not exactly the same at the different concentrations. Apart from these deviations, the general picture is the same: At high volume fractions, the correlation peak between microemulsion droplets is well defined, and forward scattering is decreased in the pure system. As before, the radius deduced from the peak position ($q_o$ = 0.0237 Å$^{-1}$), 96 Å, is close to the Guinier radius measured at low concentration. This indicates that droplets have changed either shape or become more polydisperse, without changes in volume. As the polymer is added, the peak shifts to 0.025 Å$^{-1}$ (r = 3 and r = 10), indicating again that the local network structure is different from the one of a hard sphere fluid. At small angles, an upturn in intensity is also found. It increases for moderate connectivities (r = 3), and is suppressed at higher r values (r = 10). This low-q upturn reflects the higher compressibility of the network. It appears to be correlated with the lower shear modulus measured in the rheological experiments at connectivities close to the percolation threshold $r_c$. Only upon addition of further bridging polymers the system stiffens and large-scale fluctuations are suppressed.

**III.5 Dispersion of silica in the microemulsion**

When one incorporates the silica in the microemulsion, three questions on the structure arise naturally: (a) Is the colloidal silica well dispersed in the microemulsion ? (b) Is the structure of the microemulsion affected by the silica ? And (c), are there interactions between the silica and the microemulsion droplets ? Small angle neutron scattering is a technique for structural analysis on the nanoscale which is particularly well suited to answer these questions, due to the possibility to render components visible or invisible by isotopic substitution, as done for microemulsions before [37]. Here we simply prepare samples with $D_2O$ or $D_2O/H_2O$-mixtures, under otherwise identical conditions. This allows us to contrast match either the silica, or the swollen micelles or droplets (cf. table 1 for scattering length densities), thereby answering questions (a) and (b). The last question can then be addressed by measuring cross-correlations. For the scope of this article, only the first issue will be treated, and we refer the reader to a forthcoming article on the other two.

The swollen micelles of system A have been contrast matched by setting the scattering length density of the solvent to $0.3\ 10^{10} cm^{-2}$. The coherent part of the scattered intensity is then entirely due to the silica. We start with the silica added to the pure swollen micelles, without polymer. In Figure 9, the scattered intensity is superimposed to the scattering from an independently measured pure silica dispersion at the same pH and silica volume fraction (pH 10, 6%v). Both in log-log and linear scale (not shown), the intensities are very close, indicating that the silica is dispersed among the swollen micelles in the same way as in pure water. In the inset, the same measurement performed with system B is shown, giving the same result. In this case, the solvent scattering length density is set to $-0.3\ 10^{10} cm^{-2}$.

As usual, the dispersions can be characterised through the intensity maximum. The intensities are rather peaked, indicating strong repulsion between silica beads. This has been well described with the Hansen-Hayter-Penfold structure factor (RMSA) in a very similar system [55]. From the peak position ($q_o = 0.0165$ Å$^{-1}$ at 6%v; at 0.0157 at 5%v), we can deduce the average volume of interacting objects ($\approx 3.2 \; 10^6$ Å$^3$), which is found to be comparable to the volume of single beads ($2.2 \; 10^6$ Å$^3$) [56], the minor deviation being certainly due to the rough description of the structure factor with a cubic lattic. It is concluded that the silica particles are dispersed as single beads in mutually repulsive interaction.

We now turn to the dispersion of silica in the microemulsion networks made by addition of telechelic PEO to the microemulsion. First, we have checked that addition of pure, non-modified, PEO to silica dispersions does not modify the structure of the silica by SAXS (data not shown). Secondly, scattering experiments identical to the ones without polymer have been performed. The result is shown in Figures 10a for networks of system A and in 10b for networks of system B. In the inset, the same data is plotted in linear scale, which emphasizes the (nonetheless minor) deviations caused in the silica structure by the presence of the microemulsion network.

At first sight in the log-log presentation, the intensities of the silica still superimpose reasonably well, indicating that the good dispersion is maintained. A closer look in linear scale – shown in the inset – reveals deviations which are not the same in the two systems. In system A, which is rather dilute (4%), the silica peak height decreases by some 5%, and the peak moves to the right in the microemulsion network, from $q_o = 0.0165$ Å$^{-1}$ to $0.0188$ Å$^{-1}$, presumably indicating a small decrease in repulsion. On the contrary, in the more concentrated system B (20%), the influence of the microemulsion network is more pronounced: The peak height decreases by about 25%. The peak position hardly moves (from $q_o = 0.0157$ Å$^{-1}$ to $q_o = 0.0162$ Å$^{-1}$), which means that the most probable distance remains constant, whereas the interaction is weakened. The origin of these changes in silica-silica interaction is unclear at the moment. One may speculate on a combination of adsorption of some surfactant on the silica beads [55, 57], at different cosurfactant-to-surfactant ratios, building up different types of micellar structures (bilayers, adsorbed micelles,…) on the silica surface. Here our scattering experiments with contrast-matched silica might give an answer. A second point, the quite different volume fractions of system A and B, may also have its influence. Indeed, space is far more crowded in system B, which certainly affects the silica-silica correlations. Numerical simulations of bead interactions which will hopefully advance our understanding are currently in progress in our lab.

**IV. Numerical simulation results on loops and bridges**

In this article, we have investigated experimentally two systems of model microemulsion networks filled with silica. The two systems have been chosen due to their difference in droplet size R with respect to the polymer radius of gyration $R_g$, in order to see how this influences the characteristics of the microemulsion network. Note that simulations of such networks have been published, with an effective interaction potential, empiric or based on a polymer mean-field theory [58, 59]. There are several theoretical predictions concerning the colloidal pair potential induced by a telechelic polymer [12, 39, 60]. They are usually either based on a Derjaguin approximation, or they do not take excluded volume into account, with the exception of a self-consistent field calculation focussed on adsorption [60]. The first approximation can by nature not account for, e.g., two small spheres connected by a polymer molecule of similar size, whereas the second one does not allow understanding of many chain situations. Therefore we have undertaken numerical simulations of chains connecting two colloidal particles, and first results seem to capture several interesting features, outlined below [30].

The heart of the algorithm is to set two particles of radius R (in reduced units $\alpha = R/R_g$) at a surface-to-surface distance d ($\beta = d/R_g$), and create ideal chains starting on the surface of either chain by lattice Monte Carlo simulation. $R_g$ is the radius of gyration of the free chains, $R_g = b(N/6)^{1/2}$, with b = 1 the unit length in this simulation, and the number of segments per chain N = 200, comparable to the experimental value. The induced pair potential can be calculated from the partition function, which is itself given by the sum of all thermodynamically acceptable configurations, namely bridges and loops [38]. An analogous algorithm is used for self-avoiding chains, which includes site-occupation test and rejection of the incomplete chains. As we will see from the error-bars, this considerably decreases the statistics. The reduced units $\alpha$ and $\beta$ are of course expressed in terms of the free self-avoiding $R_g$.

In a first approach, we have determined the relative proportion of bridges and loops between two beads among $16.10^6$ tries of chain construction. These tries are regrouped as 800 runs of 20'000 chain constructions each, which allows us to estimate error bars from the dispersion of the results. Possible outcomes of a chain construction are failure (a chain hitting a bead), the second sticker not hitting the bead, and loops or bridges. As a first encouraging result, the proportion of loops and bridges among the chains where both stickers are on a bead is shown in Figure 11a for two bead sizes α = 1 and 2.5. For comparison, the same result for self-avoiding chains is plotted in Figure 11b ($2.10^5$ tries).

It is clear from Figure 11, that there are roughly between two and four times more loops than bridges at small interbead distances (β < 1, i.e. d comparable to the radius of gyration). At larger distances, the fraction of bridges tends to zero, as expected. If we compare the proportions for different bead sizes, it appears that the smaller beads have a higher number of bridges, suggesting the formation of a stronger network if the bead concentrations are adapted such that the interbead distance is comparable to $R_g$, as done in our experiments. The fraction of 20-30% of bridges seems to be the correct order of magnitude, if we refer to the moduli of the microemulsion networks discussed in section III.2, i.e. 1000 Pa instead of 3000 Pa for system A, and 250 Pa instead of 900 Pa for system B. Another interesting point concerns the evolution of the proportion of bridges with distance β. Figure 11 suggests that the number of bridges is approximately constant up to the radius of gyration (β = 1), and then decays between β = 1 and 2. This seems to be related to the typical chain end-to-end distance, which is about a factor 2 to 3 (√6) bigger than the radius of gyration [61]. This justifies a posteriori our choice in bead concentration, fixed such that the average surface-to-surface distance is between one and two $R_g$. The comparison of Figures 11a and b shows us that the same observations are true for real chains, at least at low polymer amount. As with ideal chains, only one real chain is simulated at a time, and we now plan to extend the calculation to more self-avoiding chains on one bead.

To finish, some fundamental differences between our simple simulation and the experimental system should not be ignored. First of all, the experimental system is a many particle system, where chains can form bridges between neighbouring beads in all directions, and not only between one given pair. Secondly, the experimental connectivities r are relatively close to the percolation threshold, which is why a mean field, homogeneous distribution of links does not describe reality, as already discussed above.

**V. Concluding remarks**

The structure and rheology of microemulsion networks made of microemulsion droplets connected by telechelic polymer has been investigated. We have set up a model system in which the size-ratio between microemulsion droplet and the polymer coil can be varied continuously by adapting the size of the droplets. Experiments with polymer molecules of different masses are planned in the near future. The idea of these experiments is backed up by the results of numerical simulations on the proportion of loops and bridges between colloidal particles. The latter are found to vary with the aforementioned size-ratio, and also show quantitatively that the bridging interaction decays strongly at surface-to-surface distances between the radius of gyration and the typical end-to-end distance. The structural characterisation of the droplets and networks has been performed by small angle scattering. Using a Reverse Monte Carlo technique the increase of repulsive interactions induced by the polymer chains has been evidenced. The rheological properties of the networks have been measured and interpreted in terms of simple percolation expressions. The critical connectivity is found to be 2.5 and 3.5, for the two systems, a number which can be compared to the one predicted by numerical simulations on cubic networks, 1.5 [62]. If we use the estimation of the proportion of bridges obtained by our own numerical simulation presented here, 20-30%, we observe that given the differences in structure (liquid vs. cubic), the agreement is reasonable.

Finally, a first approach to microemulsion networks filled with silica nanoparticles has been presented. We have characterised the rheological response of these 'nanocomposites', and found a particularly strong reinforcement factor, well above purely hydrodynamic reinforcement. In parallel, the good dispersion of the nanosilica beads has been shown to be hardly modified by Small Angle Neutron Scattering using isotopic substitution. This finding rules out any structural contribution to the reinforcement, like formation of a silica network or large interacting aggregates. We believe that these are strong indications on a reorganisation of the network around the nanoparticles, causing a considerable increase of the number of rheologically active bridges. It is planned to present the corresponding data in a forthcoming article.

**Acknowledgements:** Work conducted within the scientific program of the European Network of Excellence *Softcomp*: 'Soft Matter Composites: an approach to nanoscale functional materials', supported by the European Commission. Silica stock solutions were a gift from Akzo Nobel.

**Appendix: Reverse Monte Carlo Modelling**

The aim of our Reverse Monte Carlo algorithm [49, 50] is to find a distribution of spheres in a box whose scattering, i.e. experimental micellar form factor times the structure factor, is compatible with the experimental data. In the beginning, 8000 particles are randomly distributed in a cubic box the size of which is fixed by the volume fraction, here 4%. It is checked that the structure factor of this initial dispersion is unity in the q-range of interest. Then spheres are moved individually by random repositioning within the box. After each step, the pair distribution function and the structure factor are recalculated, and if the agreement with the experimental intensity increases, the move is accepted, otherwise rejected. In order to speed up calculations, hard sphere interaction is not enforced. Beads are 'transparent' and are placed anywhere in the box without checking collisions. The idea is that the pair correlation function consistent with the scattered intensity should reflect the hard sphere character of the beads. This is found to be mostly true, with very few exceptions, which lead to a non-zero pair correlations below r = 2R, discarded in the plot.

**Table 1:**

|  | $\Omega$ | $\Gamma$ | $\Phi$ | R (Å) Guinier | Scattering length density $\rho$ ($10^{10}$ cm$^{-2}$) |
|---|---|---|---|---|---|
| System A (swollen micelles) | 0 | 0.05 | 4% | 32 | 0.2 |
| System B (microemulsion droplets) | 0.4 | 0.7 | 20% | 104 | -0.3 |

**Figure Captions:**

**Figure 1:** Phase diagram cut of the pure microemulsion (TX-100/octanol/decane in $H_2O$) at fixed total mass fraction of 5.0%w as a function of oil-to-surfactant ratio $\Gamma$ and cosurfactant-to-surfactant ratio $\Omega$. $L_\alpha$ denotes a lamellar phase. The two samples discussed in this paper, denoted A and B, are microemulsion droplets found in the $L_1$ phase (the 'tunnel').

**Figure 2:** Radius measured in the Guinier-domain of various droplets at high dilution along the 'tunnel' of the phase diagram, as a function of the oil-to-surfactant ratio $\Gamma$. The corresponding $\Omega$-values are given in the plot. In the inset, intensities for systems A and B are shown.

**Figure 3:** Modulus G in Pa of microemulsion network (system A) as a function of network connectivity r. The continuous line is a fit with a percolation model.

**Figure 4:** Modulus G in Pa of microemulsion network (system B) as a function of network connectivity r. The continuous line is a fit with a percolation model.

**Figure 5:** Storage ($\bullet$) and loss ($\square$) modulus as a function of $\omega$ of (a) a microemulsion network (system A, r = 10) and (b) a filled microemulsion network (system A, r = 10, $\Phi_{si}$ = 5.7%). With silica, departures from Maxwellian behaviour are emphasized in G'' at high $\omega$, and in G' at low $\omega$.

**Figure 6:** Modulus G in Pa of filled microemulsion networks (system A, r = 5($\bullet$) and 10 ($\square$)) for increasing silica volume fraction. (b) Reinforcement factor representation $G/G(\Phi_{si} = 0)$ of the same quantity.

**Figure 7:** (a) Modulus G of filled microemulsion networks (system B, r = 10($\bullet$), 15($\blacksquare$) and 20($\blacktriangle$)) for increasing silica volume fraction. (b) Reinforcement factor representation $G/G(\Phi_{si} = 0)$ of the same quantity.

**Figure 8:** Scattered intensity I as a function of wave vector q for system A in pure $D_2O$. (□) Pure swollen micelles at 1% rescaled to $\Phi = 4\%$. (○) Pure swollen micelles at $\Phi = 4\%$. (●) Swollen micelles with added polymer ($\Phi = 4\%$, r = 5). In the inset, the pair correlation function extracted by RMC is shown: (○) r = 0, $\Phi = 4\%$. (▲) r = 3, $\Phi = 4\%$ (●) r = 5, $\Phi = 4\%$.

**Figure 9:** Comparison of the scattered intensity of pure silica in water (●, $\Phi_{si} = 6\%$) to the scattering from silica among contrast matched swollen micelles (○, $\Phi = 4\%$, $\Phi_{si} = 6\%$, system A, no polymer, rescaled to identical contrast conditions). In the inset, the same comparison for system B (● = pure silica, ○ = silica in network, $\Phi = 20\%$, $\Phi_{si} = 5\%$, no polymer).

**Figure 10:** (a) Comparison of the scattered intensity of pure silica in water ((●), $\Phi_{si} = 6\%$) to the scattering from silica among contrast-matched swollen micelle network ((○), $\Phi = 4\%$, $\Phi_{si} = 6\%$, system A, r = 10, rescaled to identical contrast conditions). In the inset the same data in linear representation. (b) The same comparison for system B (● = pure silica, ○ = silica in network, $\Phi = 20\%$, $\Phi_{si} = 5\%$, r = 10). In the inset the same data in linear representation.

**Figure 11:** Simulation result of the proportion (in %) of loops and bridges for two bead sizes ($\alpha = 1$ and 2.5), as a function of reduced surface-to-surface distance $\beta$. (a) Ideal chains (N = 200), (b) Self-avoiding chains (N = 100).

**Figures**

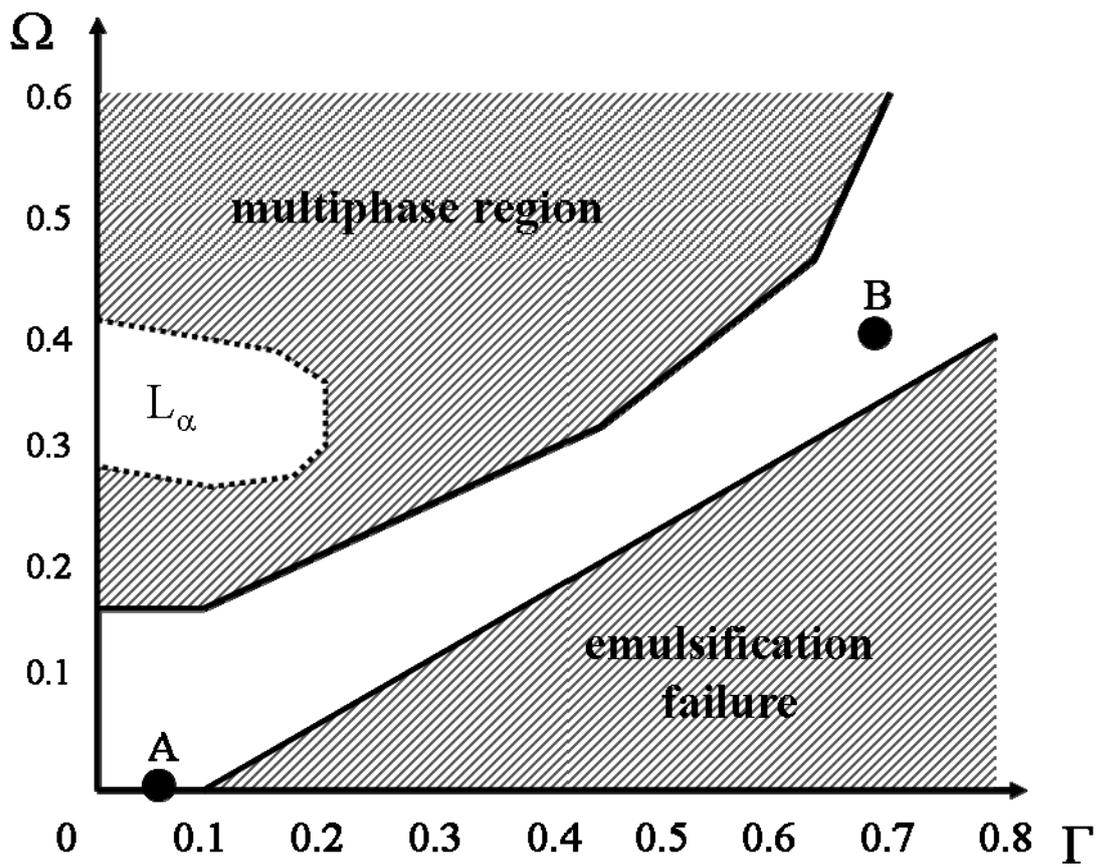

Figure 1 (Puech et al)

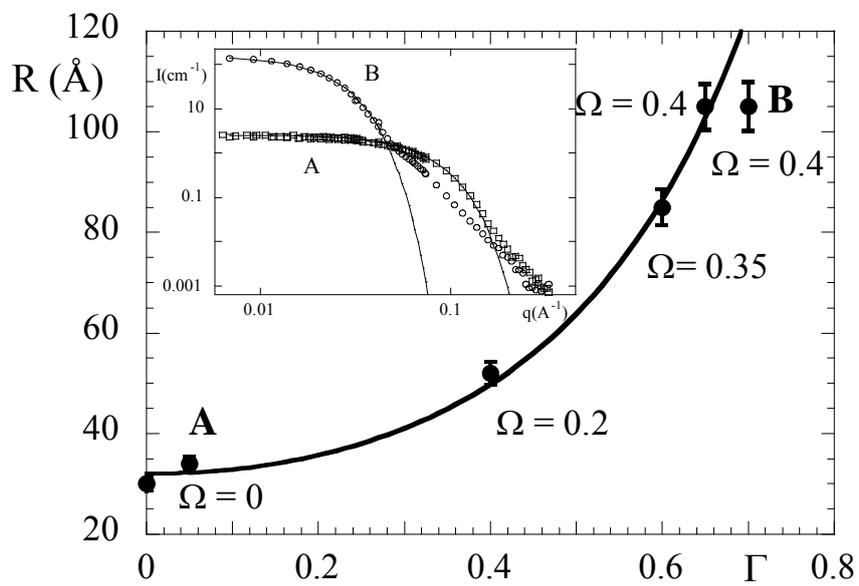

Figure 2 (Puech et al)

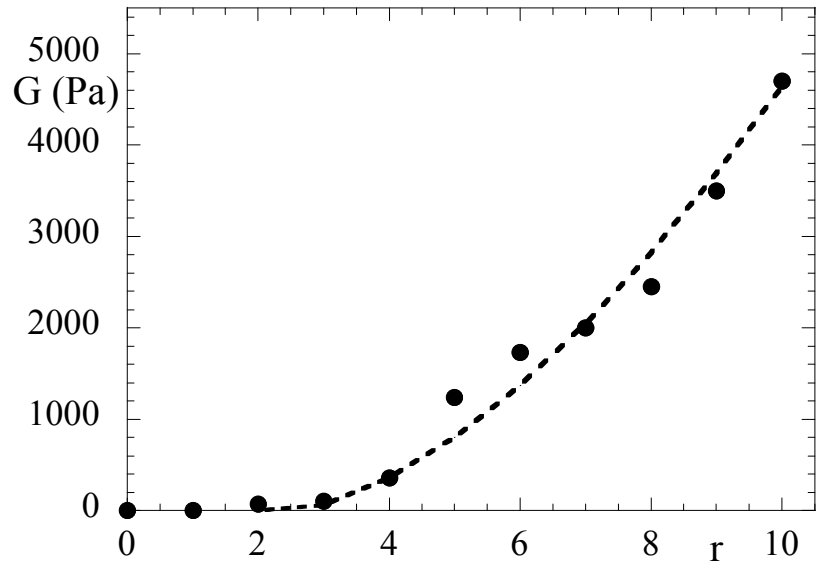

Figure 3 (Puech et al)

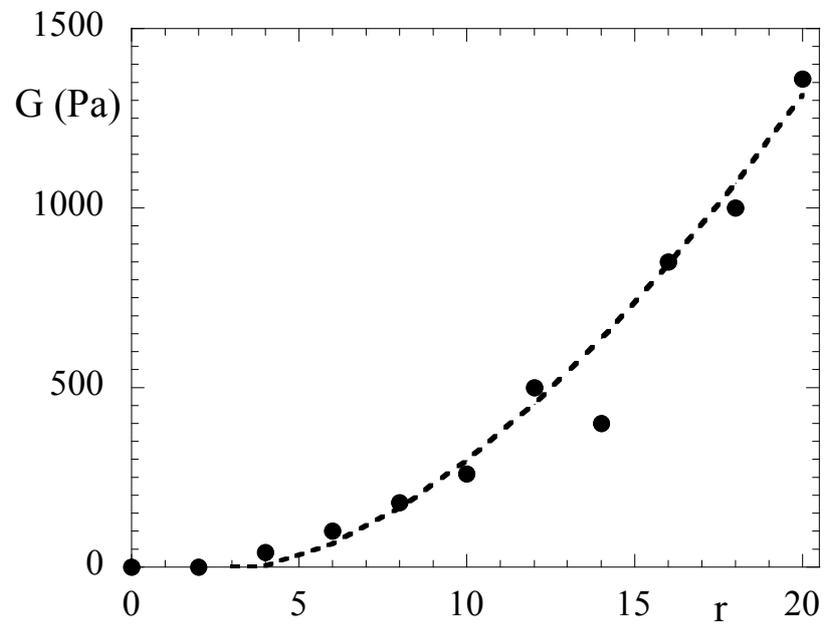

Figure 4 (Puech et al)

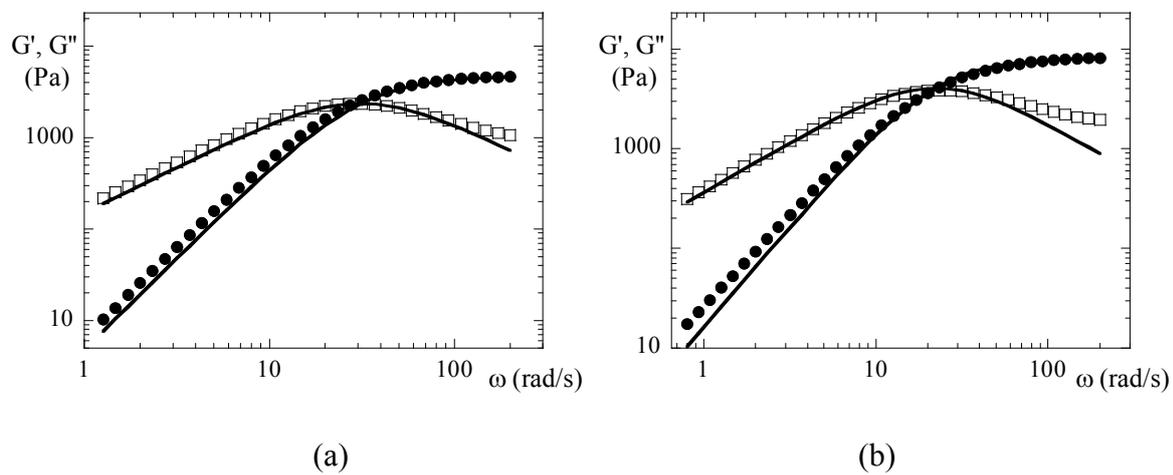

(a)  (b)

Figure 5 (Puech et al)

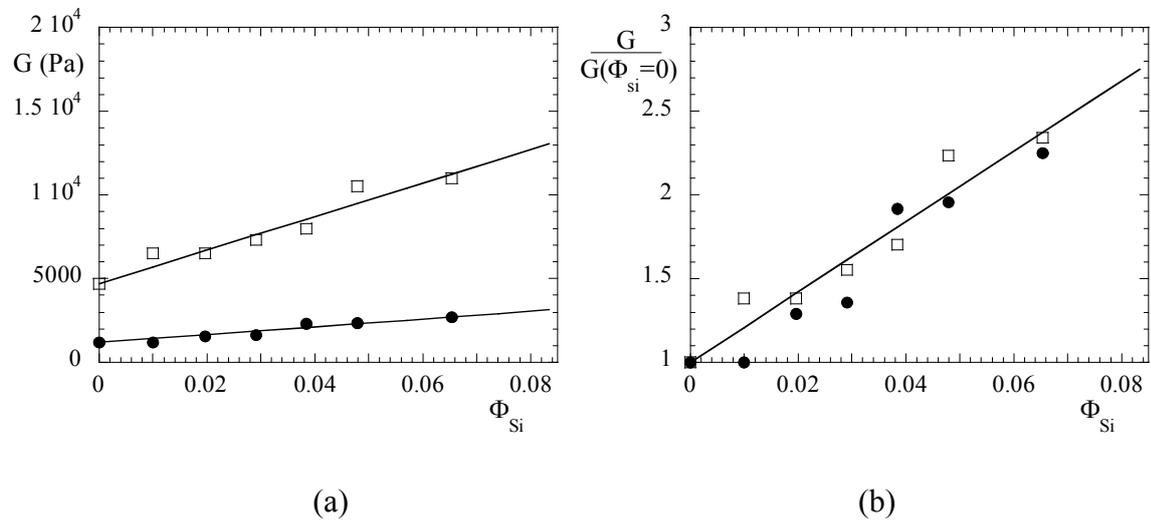

(a)                                      (b)

Figure 6 (Puech et al)

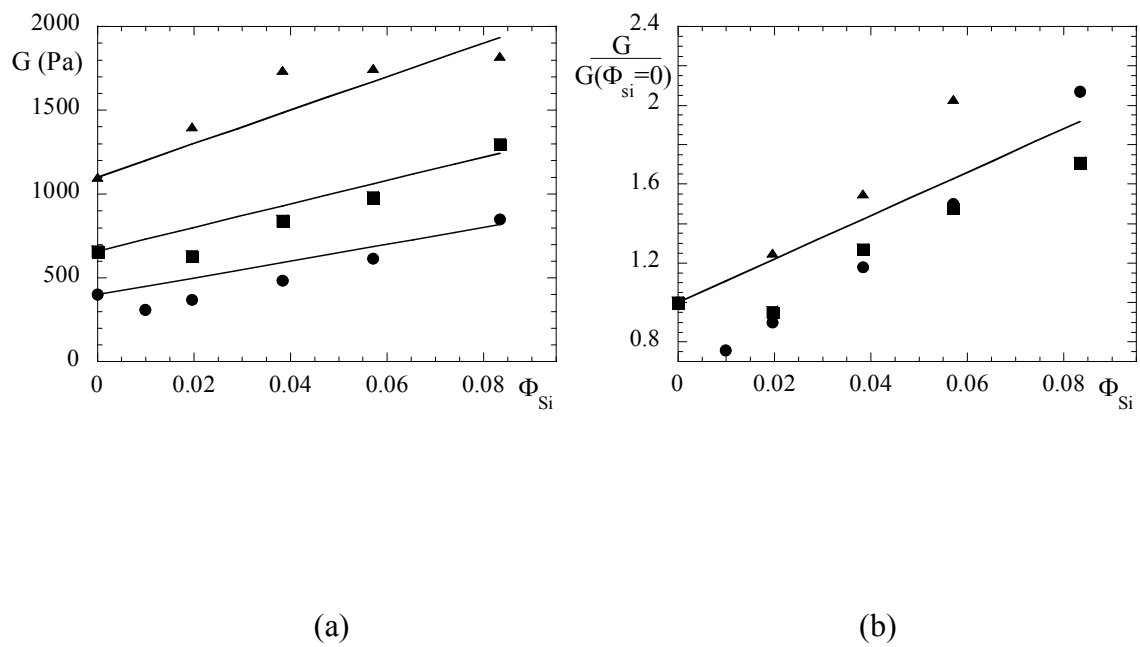

(a) (b)

Figure 7 (Puech et al)

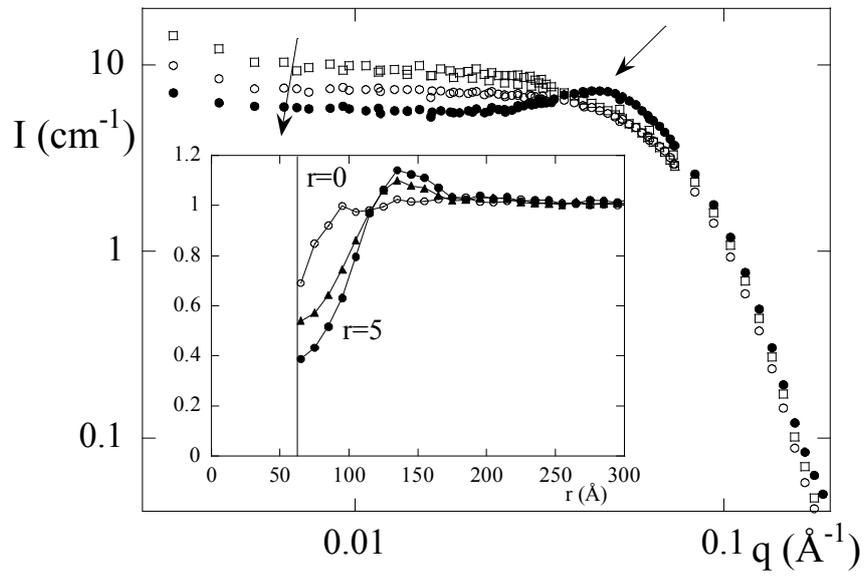

Figure 8 (Puech et al)

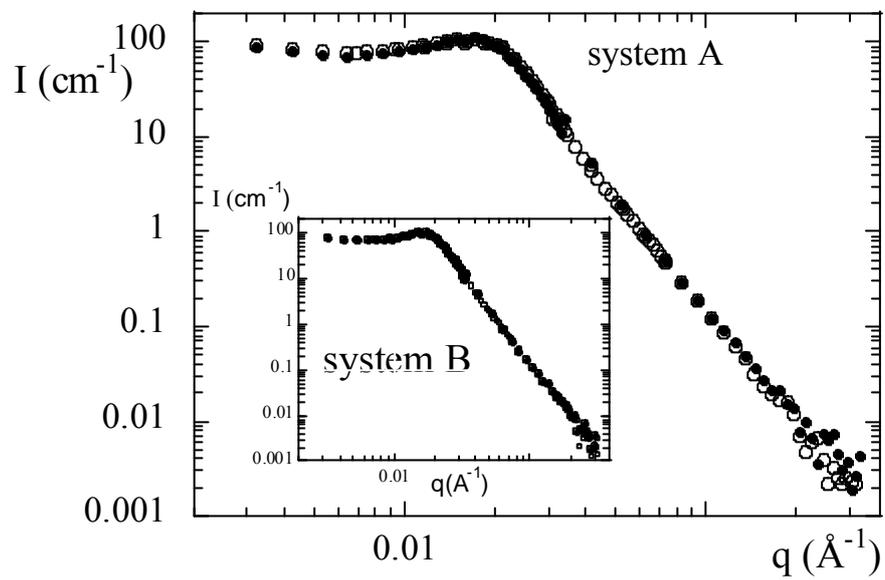

Figure 9 (Puech et al)

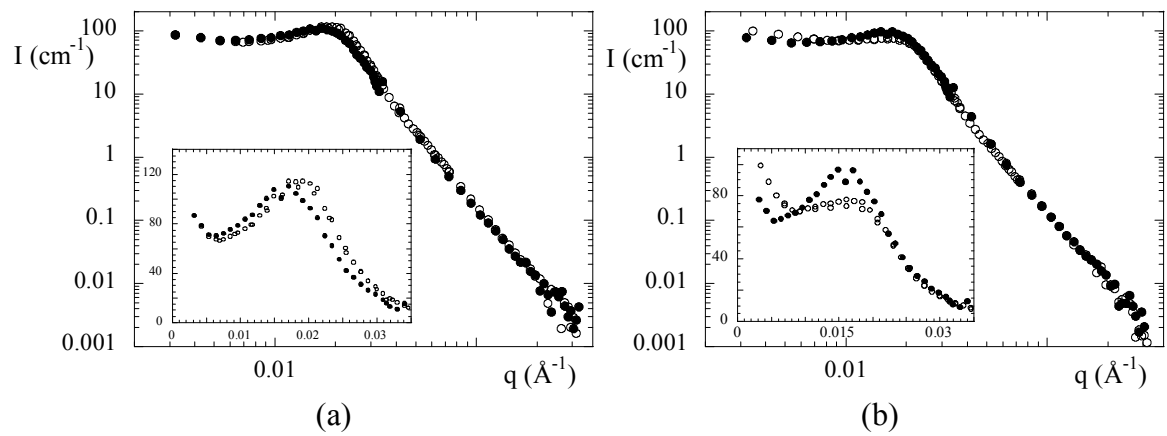

Figure 10 (Puech et al)

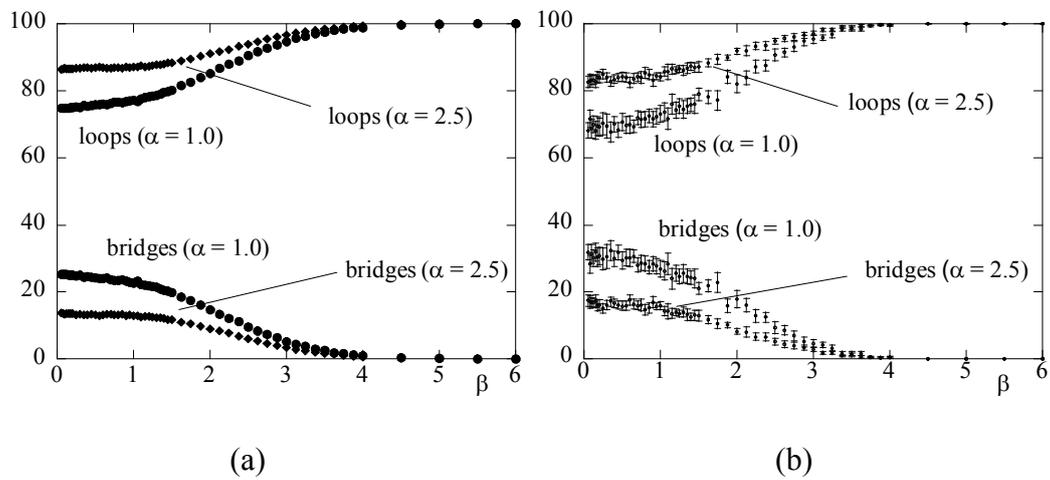

(a)                      (b)

Figure 11 (Puech et al)